\documentclass[aps,prl,twocolumn,superscriptaddress,showpacs]{revtex4}
\usepackage{graphicx}
\usepackage{calc}
\usepackage{bm}

\begin{document}

\title{Evidence for the Luttigger liquid density of states in transport across the incompressible stripe
at fractional filling factors}

\author{E.V.~Deviatov}
\email[Corresponding author. E-mail:~]{dev@issp.ac.ru}
 \affiliation{Institute of Solid State
Physics RAS, Chernogolovka, Moscow District 142432, Russia}

\author{A.A.~Kapustin}
\affiliation{Institute of Solid State Physics RAS, Chernogolovka,
Moscow District 142432, Russia}

\author{V.T.~Dolgopolov}
\affiliation{Institute of Solid State Physics RAS, Chernogolovka,
Moscow District 142432, Russia}

\author{A.~Lorke}
\affiliation{Laboratorium f\"ur Festk\"orperphysik, Universit\"at
Duisburg-Essen, Lotharstr. 1, D-47048 Duisburg, Germany}

\author{D.~Reuter}
\affiliation{Lehrstuhl f\"ur Angewandte Festk\"orperphysik,
Ruhr-Universit\"at Bochum, Universit\"atsstrasse 150, D-44780
Bochum, Germany}

\author{A.D.~Wieck}
\affiliation{Lehrstuhl f\"ur Angewandte Festk\"orperphysik,
Ruhr-Universit\"at Bochum, Universit\"atsstrasse 150, D-44780
Bochum, Germany}

\date{\today}

\begin{abstract}
We experimentally investigate transport across the incompressible
stripe at the sample edge in the fractional quantum Hall effect
regime at bulk filling factors $\nu=2/3$ and $\nu=2/5$. We obtain
the dependence of the equilibration length, that is a
phenomenological characteristics of the transport, on the voltage
imbalance and the temperature, at high voltage imbalances.  These
dependencies are found to be of the power-law form, which is a strong evidence for the Luttigger liquid
density of states.
\end{abstract}
\pacs{73.40.Qv  71.30.+h}
\maketitle

Edge states~\cite{halperin} (ES) in the integer quantum Hall
effect (IQHE) regime are arising at the intersections of the Fermi
level and Landau levels, bent up by the edge potential.
The Landauer-Buttiker formalism~\cite{buttiker}, which is based on the
concept of one-dimensional current-carrying channels, was proposed
to calculate electron transport through ES. The concept of ES and Buttiker formulas can
also be generalised to the transport along the edge in the
fractional quantum Hall effect (FQHE)
regime~\cite{Beenakker,macdonald}. Details of the inter-ES scattering, however, depend on the edge
structure of the two-dimensional electron liquid.

In the IQHE
regime the sample edge can be regarded as \textit{smooth} - the
edge potential variation  is smaller than the cyclotron gap on the
scale of the magnetic length. The smooth sample edge is a set of
the stripes of compressible and incompressible electron
liquid~\cite{shklovsky}. This picture  was confirmed in imaging
experiments~\cite{image} and in experiments on transport across
the sample edge in quasi-Corbino geometry~\cite{alida}. The latter
ones, both as the experiment~\cite{mueller}, demonstrate that the
inter-ES scattering is mostly determined by the energetic
structure at the sample edge.

It is a common belief that FQHE is determined by the
electron-electron interactions~\cite{obzor}. Also, because of
smaller spectrum gaps, a sharp edge can be realized. ES in this
case are regarded as the realization of the one-dimensional
electron liquid, also known as chiral Luttinger liquid~\cite{wen}.
A characteristic feature of the Luttinger liquid is the gapless
collective excitation spectrum~\cite{wen,kane},  that is
determined by the hierarchical structure of the fractional ground
state~\cite{macdonald}.  The excitation spectrum defines the
tunnelling density of states, which takes a form $D(E)\sim
E^{1/g-1}$, where $g$ is of universal value $g=1/\nu$.
Experimental investigations were made in the cleaved-edge
overgrowth technique~\cite{chang,grayson}, that is the best
candidate for the sharp sample edge realization. They demonstrate
power-law $I-V$ curves for tunnelling into the fractional edge
with exponents $g$ that behave as $1/\nu$ for different $\nu$ from
1/4 to 1.

It is clear, that the gate-defined electrostatic edge can not be
regarded as  sharp, even for FQHE. Like in the integer
case~\cite{shklovsky}, the sample edge is a set of compressible
and incompressible electron liquid~\cite{Beenakker}. It was shown
in numerical calculations~\cite{chamon}, that the structure of the
compressible/incompressible stripes arises at the edge width of
the order of 5-6 magnetic lengths, and, therefore, is present for
most of real potentials. Co- and counter- propagating excitation
branches exist at the edges of the incompressible
stripes~\cite{chamon,vignale}, leading to the edge excitation
spectrum resembling neutral acoustic modes~\cite{aleiner}.
Transport between two smooth fractional edges was tested in
experiments~\cite{pellegr} in quantum point contacts (QPC), where
the power-low $I-V$ curves were reported and the influence of the
potential profile in QPC on the $g$ values~\cite{vignale} was
demonstrated.

There is, however, a possibility to investigate effects of the
neutral edge modes directly.  It is the investigation of transport
across a single incompressible stripe at one sample edge. That
allows to remove the influence of the exact form of the potential
profile, leaving the effects of the neutral excitation modes to
prevail in the tunnel density of states. Investigations could be
performed in the quasi-Corbino sample geometry~\cite{alida}, which
has the following advantages: (i) etched mesa edge allows to
create the potential profile of the intermediate strength, where
small widths of the stripes simplify neutral edge modes
excitation; (ii) split-gate with well-defined stripe structure
allows to separately contact stripes in the gate-gap; (iii)
Corbino topology provides direct measurements of the electron
transport across the incompressible stripes. This situation, up to
our knowledge, was not in focus of both experimental and
theoretical investigations before.

Here we experimentally investigate transport across the
incompressible stripe at the sample edge in the fractional quantum
Hall effect regime at bulk filling factors $\nu=2/3$ and
$\nu=2/5$. We obtain the dependence of the equilibration length,
that is a phenomenological characteristics of the transport, on
the voltage imbalance and the temperature, at high voltage
imbalances. These dependencies are found to be of the power-law form, which is a strong evidence for the
Luttigger liquid density of states.

Our samples are fabricated from  molecular beam epitaxial-grown
GaAs/AlGaAs heterostructure. It contains a 2DEG located 150~nm
below the surface. The mobility at 4K is 1.83 $\cdot
10^{6}$cm$^{2}$/Vs and the carrier density 8.49 $\cdot
10^{10}$cm$^{-2}$, as was obtained from usual magnetoresistance
measurements. Also, magnetocapacitance measurements were performed
to characterize the electron system under the gate. An interplay
between two ground states~\cite{obzor} (spin polarised (SP) at
$B=5.18$~T and spin unpolarised (SU) at $B=4.68$~T) at $\nu=2/3$
is well developed in our samples~\cite{nu23rec}  permitting the
measurements at different spin configurations of the $\nu=2/3$
ground state.

\begin{figure}
\includegraphics*[width=0.6\columnwidth]{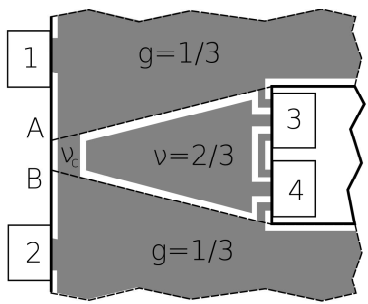}%
\caption{ Schematic diagram of the sample working area. The etched
mesa edges are shown by solid lines, the dashed lines
represent the split-gate edges. The gate-gap region at the outer mesa edge is denoted as AB. Light gray areas are the incompressible
regions at filling factors $\nu$ (in the bulk) and $g<\nu$ (under
the gate and the incompressible stripes at the mesa edges).
Compressible regions (white) are at the electrochemical potentials
of the corresponding ohmic contacts, denoted by bars with numbers.
 \label{sample}}
\end{figure}

Samples are patterned in the quasi-Corbino sample
geometry~\cite{alida}. Each sample has an etched region
inside, providing two independent mesa edges. At bulk filling
factor $\nu$ a structure of compressible and incompressible
stripes exists at every edge, see Fig.~\ref{sample}. Some of the
stripes are redirected to the other edge by depleting 2DEG under
split-gate to lower filling factor $g<\nu$. The gap in the gate at
the outer edge (the gate-gap region, denoted as AB in
Fig~\ref{sample}) is much more narrow than at the inner one, where
it is of macroscopical width $\sim 1$mm. As a result, applying a
voltage between ohmic contacts at outer and inner edges leads to
the electrochemical potential imbalance across the incompressible
stripe at local filling factor $\nu_c=g$ at the outer edge, see
Fig.~\ref{sample}. The imbalance is a maximum $V$ at the
"injection" side of the gate-gap (depends on the magnetic field
direction; for example, "A" in Fig.~\ref{sample}). Because of the
gradual equilibration across the incompressible strip, the
imbalance decreases while going from "A" to "B". The full current
$I$, transferred across the incompressible stripe can easily be
written as
\begin{equation}
I= (V/R_{eq}) (1-exp(-L_{AB}/l_{eq})),\label{eq1}
\end{equation}
where $l_{eq}$ is the phenomenological equilibration length, $R_{eq}$ is the
equilibrium Buttiker~\cite{buttiker,Beenakker} resistance. Thus, $l_{eq}$ can be obtained from experimental
$I-V$ curves with reasonable accuracy for $L_{AB}<<l_{eq}$ only. For the present experiment the gate-gap
region is extremely narrow, $L_{AB}=0.5 \mu$m, so we can expect $L_{AB}<<l_{eq}$ for any integer and
fractional fillings~\cite{mueller}.

\begin{figure}
\includegraphics[width= 0.8\columnwidth]{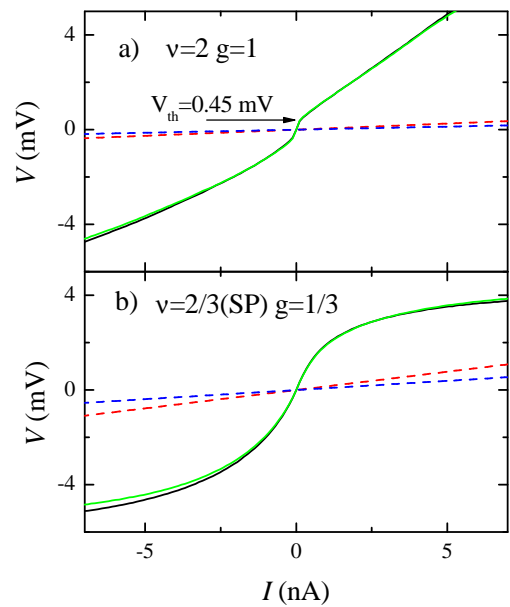}%
\caption{(Color online) $I-V$ curves for integer filling factors $\nu=2, g=1$ (a) and fractional ones
$\nu=2/3, g=1/3$ (b)
for two different contact configurations. Equilibrium lines (with $R_{eq}=2;1;6;3 h/e^2$) are  shown.
Magnetic field $B$ equals to 1.67~T for integer fillings and to 5.18~T for fractional
ones.
\label{IV}}
\end{figure}
We study $I-V$ curves in 4-point configuration, by applying
\textit{dc} current between a pair of inner and outer contacts and
measuring \textit{dc} voltage between another pair of inner and
outer contacts. This procedure is used to remove a possible
influence of the ohmic contacts on the $I-V$ curves. However, we
pay a special attention to this point: several pairs of contacts
at inner and outer edges of the sample allow us to study contact
behaviour by 2-point magnetoresistance measurements
(Corbino-contacts and non-linear ones can easily be excluded in
this way). We used 4 samples with different contact recipes
(resistance at low temperature is in the range 100-500~Ohm) and
obtain similar results. The form of $I-V$ curves was additionally
verified in 2-point measurements by applying dc voltage and
measuring dc current. The measurements are performed in a dilution
refrigerator with base temperature 30~mK, equipped with a
superconducting solenoid. The results, presented here, are
independent of the cooling cycle.

Examples of $I-V$ curves  at low (30~mK) temperature for integer
($\nu=2,g=1$) and  fractional ($\nu=2/3 (SP), g=1/3$)  filling
factors are shown in Fig~\ref{IV}, (a) and (b) correspondingly.
They are strongly non-linear, but the \textit{qualitative}
difference is obvious: (i) for integer fillings $I-V$ is
asymmetric; the positive branch of the $I-V$ trace is linear in a
wide voltage range (5~mV) and starts from the finite threshold
voltage $V_{th}=0.45$~mV; (ii) for fractional fillings $I-V$
traces are practically symmetric; both branches are strongly
non-linear and continuously start from the zero. This behaviour is
characteristic for $I-V$ curves at all integer and fractional
fillings, as investigated here as well reported
before~\cite{alida,relax,axel}. Also, a pronounced  hysteresis is
present for $\nu=2, g=1$ (not shown here), that is usual for
transport between spin-split integer ES~\cite{relax}, but no
hysteresis can be found for fractional fillings even for different
spin polarisations of the $\nu=2/3$ ground state.

$I-V$ curves are presented in Fig.~\ref{IV} for two different
contacts configurations, the difference is negligible. Thus, $I-V$
curves do reflect transport in the gate-gap across the
incompressible stripe, but not the contact configuration and
behaviour. Also, lines with calculated equilibrium Buttiker
slopes~\cite{buttiker,Beenakker} $R_{eq}=h/e^2 \nu/g(\nu-g)$ and
$R_{eq}=h/e^2/(\nu-g)$ are shown for these contact configurations
in Fig.~\ref{IV}. Experimental $I-V$ traces are significantly
above the equilibrium lines in Fig.~\ref{IV}, indicating
$L_{AB}<<l_{eq}$ regime in accordance with Eq.~\ref{eq1}.

\begin{figure}
\includegraphics[width= 0.8\columnwidth]{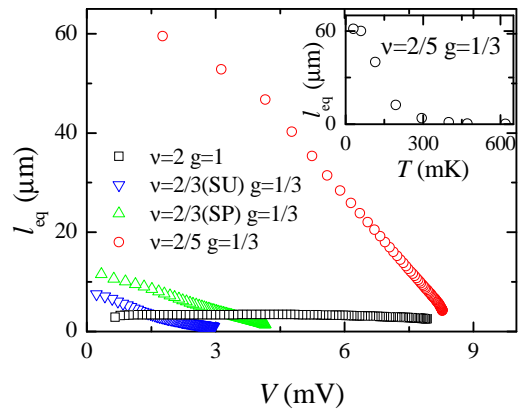}%
\caption{(Color online) Calculated $l_{eq}$ for different filling factor combinations as function of the
voltage
 imbalance across the incompressible stripe with corresponding $\nu_c=g$. Inset shows the temperature
 dependence of the $l_{eq}$ for $\nu=2/5, g=1/3$ ($B=7.69$~T) at voltage imbalance $V=1.6$~mV. \label{Leq}}
\end{figure}

In this regime, equilibration length $l_{eq}$ can be calculated from Eq.~\ref{eq1}, as shown in Fig.~\ref{Leq}
for different filling factor combinations. At integer fillings $\nu=2, g=1$, $I-V$ curve starts from the
threshold voltage $V_{th}=0.45$~mV, determined by the Zeeman splitting, because of Landau Levels bending in
the incompressible $\nu_c=g=1$ stripe. Only voltage imbalance $V-V_{th}$ can be
redistributed~\cite{alida} in Eq.~\ref{eq1} at high imbalances $V>V_{th}$ .  We find $l_{eq}$ in this
regime to be equal to $l_{eq}\sim 3 \mu$m and practically independent of $V$ in a wide (5~mV) voltage
range. This finding becomes possible only because of small $L_{AB}=0.5 \mu$m$ <<l_{eq}$, in comparison with
previous reports~\cite{alida} that were in the $l_{eq}<L_{AB}$ regime.

Just from the form of the $I-V$ curves for the fractional filling
factors we can expect a significant dependence of the
equilibration length on the voltage imbalance. This is
demonstrated in Fig.~\ref{Leq}, where $l_{eq} (V)$ are
monotonically falling functions for fractional fillings. The
results for $\nu=2/3, g=1/3$ are presented for two different spin
orientations of the $\nu=2/3$ ground state~\cite{obzor}. Both
dependencies $l_{eq}(V)$ run practically in parallel, that can be
a sign, as well as the absence of the hysteresis on $I-V$ curves,
that $\nu=2/3$ ground state polarisation is not important. The
difference in value can be ascribed to the higher $\nu_c=1/3$
incompressible stripe width in the higher  magnetic field at which
spin-polarised $\nu=2/3$ occurs. The dependence $l_{eq}(V)$ for
fractional fillings $\nu=2/5, g=1/3$ is much stronger and values
of the $l_{eq}$ are much higher. Also, the temperature dependence
of the $l_{eq}$ for $\nu=2/5, g=1/3$ is shown in the inset to
Fig.~\ref{Leq} at voltage imbalance $V=1.6$~mV. The form of the
$l_{eq}(T)$ behaviour is independent of the $V$ value.

To analyze both $l_{eq}(V)$ and $l_{eq}(T)$ dependencies, we
should mention that $l_{eq}\sim w^{-1}$, where $w$ is the
transport probability. In general, it can be written as $w \sim
D(V,T) T_0(V)$  with $D$ denoting the tunnel density of states,
and $T_0$ describing one-particle transmittance of the barrier.
$T_0$ is determined by the potential jump in the incompressible
region $\Delta$. At integer fillings $\Delta$ itself depends on
the voltage imbalance $V$, because the Landau levels bending is
diminishing~\cite{alida} at $V>0$ in the incompressible stripe. At
$V=V_{th}$ the flat-band situation $\Delta=0$ occurs and $w$ is
not sensitive to the tunnel density of states $D$, that can be
seen from the constant $l_{eq}$ in a wide voltage range. At $V<0$
no flat-band situation could exist, but $D$ is the pre-exponent
factor in $w$ that hardly be extracted. Thus, one-particle $T_0$
dominates in transport in IQHE regime.

\begin{figure}
\includegraphics[width= 0.8\columnwidth]{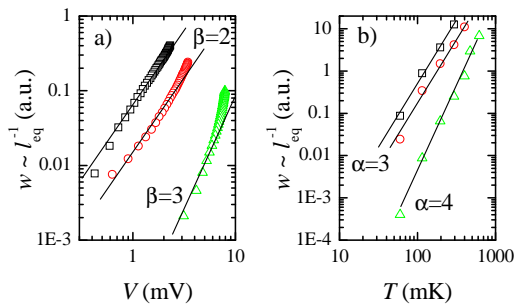}%
\caption{(Color online) $w \sim l_{eq}^{-1}$ is shown as a
function of the temperature at $V=1.6$~mV (a) and as a function of
the voltage imbalance at $T=30$~mK (b) in logarithmic scales. The
filling factors are $\nu=2/3(SU), g=1/3$ (squares);
$\nu=2/3(SP),g=1/3$ (circles); $\nu=2/5,g=1/3$ (triangles).
\label{LogLog}}
\end{figure}
For fractional filling factors we can not expect the flat-band
situation at any sign of $V$, because there are no bending of
Landau levels in this case~\cite{Beenakker}, which is the reason
to much more symmetric $I-V$ curves. However, the significantly
smaller fractional gaps~\cite{obzor} lead to $\Delta<<V$ in our
voltage range. The transmittance $T_0$ can be written as
$exp(-C\Delta^{3/2}/V)$ in the quasiclassical triangular barrier
approximation with constant $C$. In our regime $\Delta<<V$ the
tunnel barrier is so thin~\cite{triangular}, that $T_0(V)\sim 1$
has no influence on the transport probability $w$. Thus, $w$ is
mostly determined by the tunnel density of states $D$. In
Fig.~\ref{LogLog} $w \sim l_{eq}^{-1}$ is shown as a function of
the voltage imbalance $V$ (a) and the temperature (b) for
different fractional filling factors in logarithmic scales.  The
dependencies are of a power-law behavior: $w\sim V^\beta$ with
$\beta=2$ for $\nu=2/3, g=1/3$ (both spin polarizations) and
$\beta=3$ for $\nu=2/5, g=1/3$; $w(T) \sim T^\alpha$  with
$\alpha=3$ for $\nu=2/3, g=1/3$ and $\alpha=4$ for $\nu=2/5,
g=1/3$. (The last fact, that the temperature behaviour is not
activated one was also checked by plotting $w(T)$ in the Arrhenius
scales.) They are the power-law dependencies $w(V)\sim V^\beta$
and $w(T)\sim T^\alpha$ with $\beta=\alpha-1$ that were predicted
for tunnel density of states $D$, determined by the collective
excitations~\cite{wen,kane,vignale}. This conclusion is also
supported by the $w(V,T)$ independence of the $\nu=2/3$ spin
polarization (see Figs.~\ref{Leq},\ref{LogLog}), because it is
$T_0$ that is sensitive to the spin polarisation of the ground
state~\cite{relax} but not $D$. Thus, neutral excitation modes~\cite{aleiner} do
exist at the sample edge in the FQHE regime and determine
transport across the incompressible stripe at high imbalances
$V>>\Delta$.

At $\nu=2/3, g=1/3$ the neutral mode is constructed from the two excitation branches, which are
counter-propagating~\cite{chamon} along the edges of the $\nu_c=1/3$ incompressible stripe.
 At $\nu=2/5, g=1/3$ we study transport across the same $\nu_c=1/3$ incompressible stripe,
but the experimental exponents $\alpha,\beta$ are different. The edge of the $\nu=2/5$ bulk incompressible
state is extremely close to $\nu_c=1/3$ stripe in this case, because of $\nu-\nu_c<<\nu$.
Thus, we can expect some influence in $D$ also from the edge
excitations of $\nu=2/5$ bulk incompressible state, affecting the
exponents in power-low $D(V,T)$. This influence can also be responsible for the deviations from the straight
lines in the Fig.~\ref{LogLog} (a) at high voltages. It is still to be
calculated in further theoretical investigations.

We wish to thank  D.E.~Feldman for fruitful discussions and
A.A.~Shashkin for help during the experiment. We gratefully
acknowledge financial support by the RFBR, RAS, the Programme "The
State Support of Leading Scientific Schools", Deutsche
Forschungsgemeinschaft, and SPP "Quantum Hall Systems", under
grant LO 705/1-2. E.V.D. is also supported by MK-4232.2006.2 and
Russian Science Support Foundation.


\begin{thebibliography}{99}
\bibitem{halperin} B. I. Halperin, Phys.\  Rev.\ B  {\bf
25}, 2185 (1982).
\bibitem{buttiker} M. B\"uttiker, Phys. Rev. B {\bf 38}, 9375 (1988).
\bibitem{Beenakker} C. W. J. Beenakker, Phys. Rev. Lett. \textbf{64}, 216 (1990).
\bibitem{macdonald} A. H. MacDonald, Phys. Rev. Lett. \textbf{64}, 220 (1990).
\bibitem{shklovsky} D. B. Chklovskii, B. I. Shklovskii, and L. I.
Glazman, Phys. Rev. B {\bf 46}, 4026 (1992).
\bibitem{image} E. Ahlswede, J. Weiss, K. von Klitzing, K. Eberl, Physica E \textbf{12}, 105 (2002).
\bibitem{alida} A. W\"urtz, R. Wildfeuer, A. Lorke,
E. V. Deviatov, and V. T. Dolgopolov, Phys. Rev. B {\bf 65},
075303 (2002); E. V. Deviatov, V. T. Dolgopolov, A. Wurtz, JETP
Lett. {\bf 79},  618 (2004).
\bibitem{mueller} G. M\"uller, D. Weiss, A. V. Khaetskii,
 K. von Klitzing,  S. Koch, H. Nickel, W. Schlapp, and R. L\"osch,
 Phys. Rev. B {\bf 45}, 3932 (1992).
\bibitem{obzor} For a review on the FQHE, see T. Chakraborty, Adv. Phys. \textbf{49}, 959
(2000).
\bibitem{wen} Xiao-Gang Wen, Phys. Rev. B \textbf{43}, 11025 (1991); Phys. Rev. Lett. \textbf{64}, 2206
(1990); Phys. Rev. B \textbf{44}, 5708 (1991).
\bibitem{kane} C.L. Kane and M.P.A. Fisher, Phys. Rev. B \textbf{46}, 15233 (1992); Phys. Rev.
Lett. \textbf{68}, 1220 (1992).
\bibitem{chang} A. M. Chang, L. N. Pfeiffer, and K. W. West; Phys. Rev. Lett. \textbf{77}, 2538
(1996).
\bibitem{grayson}M. Grayson, D. C. Tsui, L. N. Pfeiffer, K. W. West, and A. M.
Chang, Phys. Rev. Lett. \textbf{80}, 1062 (1998); M. Hilke, D. C.
Tsui, M. Grayson, L. N. Pfeiffer, and K. W. West, Phys. Rev. Lett.
\textbf{87}, 186806 (2001).
\bibitem{chamon} C. d. C. Chamon and X. G. Wen, Phys. Rev. B \textbf{49},
8227 (1994).
\bibitem{vignale} S. Conti and G. Vignale, Phys. Rev. B \textbf{54},
14309 (1996).
\bibitem{aleiner} I.L. Aleiner and L.I. Glazman,  Phys. Rev. Lett.
\textbf{72}, 2935 (1994).
\bibitem{pellegr} S. Roddaro, V.
Pellegrini, F. Beltram, G. Biasiol, and L. Sorba, Phys. Rev. Lett.
\textbf{93}, 046801 (2004); S. Roddaro, V. Pellegrini, F. Beltram,
L. N. Pfeiffer, and K. W. West Phys. Rev. Lett. \textbf{95},
156804 (2005).
\bibitem{nu23rec} V.T. Dolgopolov \textit{et. al.},
cond-mat/0606716.
\bibitem{relax} E. V. Deviatov, A. Wurtz, A. Lorke, M. Yu. Melnikov, V. T. Dolgopolov, D. Reuter, A. D.
 Wieck, Phys. Rev. B \textbf{69}, 115330 (2004).
\bibitem{axel} A. Wurtz, E. V. Deviatov, A. Lorke, V. T. Dolgopolov, D. Reuter, and A. D.
 Wieck,  Physica E \textbf{22}, 177 (2004).
\bibitem{triangular} $\Delta<<V$ supports the triangular barrier approximation for any initial barrier
profile, but breaks the quasiclassical limit - one-particle tunnell probability is close to one. We need only
in the last fact here.
\end{thebibliography}
\end{document}